\documentclass[reprint,
aps,superscriptaddress,twocolumn,showpacs,longbibliography]{revtex4-2}
\usepackage{graphicx,color}
\usepackage{amsmath,amsfonts,enumerate,amsthm,amssymb,bbm}
\usepackage[colorlinks=true,citecolor=blue,linkcolor=magenta]{hyperref}
\usepackage{multirow}
\usepackage{bbold}
\usepackage{braket}
\usepackage{soul}
\usepackage[caption = false]{subfig}
\usepackage{scrextend}
\usepackage{xcolor}
\usepackage{dsfont}
\usepackage{nicefrac}
\usepackage[linesnumbered,ruled,vlined]{algorithm2e}
\usepackage{quantikz}
\usepackage{apptools}

\mathchardef\ordinarycolon\mathcode`\:
\mathcode`\:=\string"8000
\begingroup \catcode`\:=\active
  \gdef:{\mathrel{\mathop\ordinarycolon}}
\endgroup

\theoremstyle{plain}

\theoremstyle{definition}

\theoremstyle{remark}

\AtAppendix{\counterwithin{thm}{section}}
\AtAppendix{\counterwithin{corol}{section}}
\AtAppendix{\counterwithin{lem}{section}}
\AtAppendix{\counterwithin{propos}{section}}
\AtAppendix{\counterwithin{defn}{section}}
\AtAppendix{\counterwithin{rmk}{section}}
\AtAppendix{\counterwithin{ex}{section}}

\def\<{\langle}

\def\>{\rangle}
\def\<{\langle}

\begin{document}

%\title{Strong Gradient Variance in Highly Symmetric Nuclear Systems}
%\title{Strong Gradients in Highly Symmetric Nuclear Systems}
%\title{Exploiting symmetries in the Agassi model for improved trainability}

\title{Exploiting symmetries in nuclear Hamiltonians for ground state preparation}

% \title{(?) Strong Gradients for Ground State Preparation of 2-level Nuclear Hamiltonians}

\author{Joe Gibbs}
\affiliation{School of Mathematics and Physics, University of Surrey, Guildford, GU2 7XH, UK}
\affiliation{AWE, Aldermaston, Reading, RG7 4PR, UK}

\author{Zo\"e Holmes}
\affiliation{Institute of Physics, Ecole Polytechnique Fédéderale de Lausanne (EPFL), CH-1015, Lausanne, Switzerland}

\author{Paul Stevenson}
\affiliation{School of Mathematics and Physics, University of Surrey, Guildford, GU2 7XH, UK}

\date{\today}

\begin{abstract}
The Lipkin and Agassi models are simplified nuclear models that provide natural test beds for quantum simulation methods. Prior work has investigated the suitability of the Variational Quantum Eigensolver (VQE) to find the ground state of these models. There is a growing awareness that if VQE is to prove viable, we will need problem inspired ans\"atze that take into account the symmetry properties of the problem and use clever initialization strategies. Here, by focusing on the Lipkin and Agassi models, we investigate how to do this in the context of nuclear physics ground state problems. We further use our observations to discus the potential of new classical, but quantum-inspired, approaches to learning ground states in nuclear problems.
\end{abstract}

\maketitle

%\paragraph*{Introduction}

\section{Introduction}

Nuclei, as interacting systems of spin-1/2 and isospin-1/2 fermions, are natural candidates for simulation on quantum hardware. Moreover, since they are composed of an intermediate number of particles, too many for exact diagonalisation and too few for the thermodynamic approximation, they often prove challenging to simulate accurately by other methods. Nuclei are further charactersised by the presence of the strong nuclear interaction which in principle includes many-body forces. In contrast to molecules, Nuclei have a single centre but can exhibit deformation and some quasi-molecular structure.

The configuration interaction form of the nuclear shell model features the use of single-particle orbitals as basis states, truncated suitably to accommodate only low energy excitations.  The nuclear Hamiltonian, renormalised for the truncated basis, is then diagonalised to obtain the nuclear spectrum.  Depending on the truncation details, Hamiltonian dimensions can exceed $~10^{20}$ \cite{dean_computational_2008} and instead methods which target the interesting extremal eigenvalues, such as the Lanczos method, are usually employed.  Nevertheless, the method is still prohibitive for many cases of interest and alternative algorithms with more efficient scaling are called for. Exact solutions are known for simplified versions of the shell model, such as the Lipkin model~\cite{lipkin1965validity} (also known as the Lipkin-Meshkov-Glick model) and the slightly more complex Agassi model~\cite{agassi1966validity}, which we detail later in the present manuscript, and so form natural test beds for new methods.

\begin{figure*}[t!]
\centering
\includegraphics[width=0.8\textwidth]{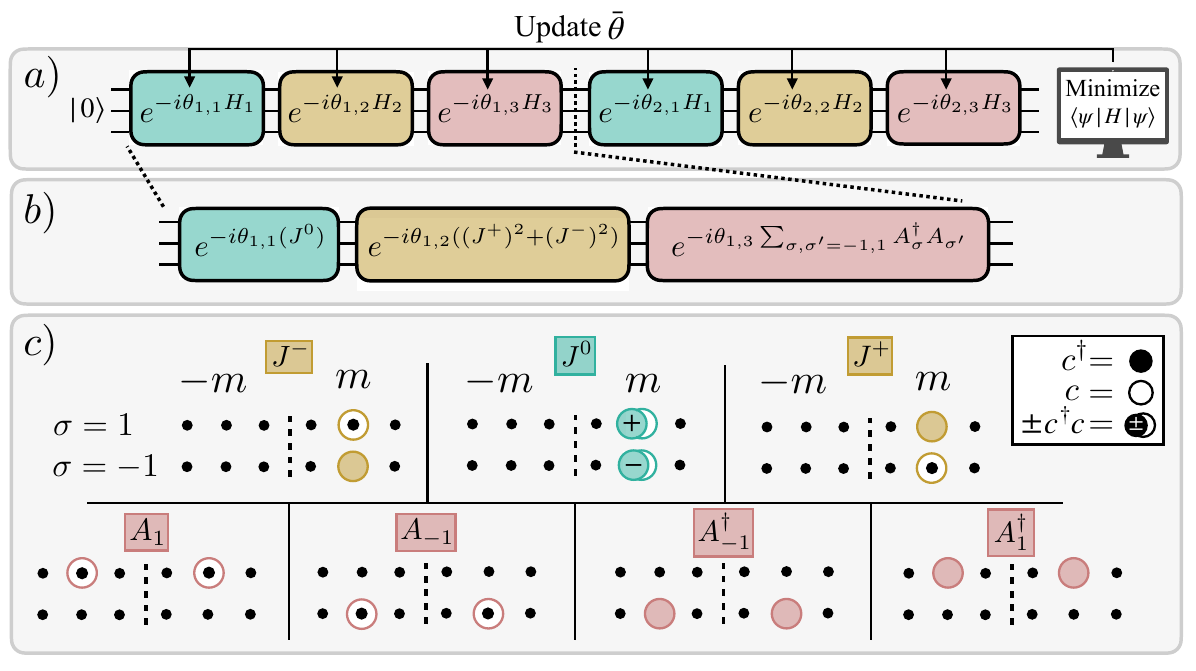}
\vspace{-2mm}
\caption{\textbf{The Hamiltonian Variational Ansatz for the Agassi Model. } a) Cartoon representation of the VQE algorithm using a Hamiltonian variational ansatz, shown with 3 gate blocks and 2 ansatz layers b) The unitaries in each layer of the ansatz are rotations with the hermitian generators corresponding to the three terms in the Agassi Model Hamiltonian Eq.~\ref{eq:agassi_ham}. (We note that in practise the gates in b) would need to be Trotterised to be implemented with a standard gate set, but we do not consider this additional source of error here.) c) A pictoral representation of the operators in the Hamiltonian. Each term is a sum over magnetisation values, and we show the creation/annihilation operators for $m=2$.}
\label{fig:FancyFig}
\vspace{-1.5em}
\end{figure*}

One recently-proposed method for finding ground states for nuclear problems is the Variational Quantum Eigensolver (VQE)~\cite{peruzzo2014variational}. The VQE algorithm uses a hybrid quantum classical optimisation loop~\cite{cerezo2020variationalreview}, where the energy of a trial quantum state with respect to a given target Hamiltonian is evaluated on a quantum computer, while a
classical optimiser trains a parameterised quantum circuit
to minimise this energy. If successfully trained, the state that minimises the energy of the Hamiltonian will be an estimate of its ground state, and the corresponding energy will be an estimate of the ground state energy. Core to the success of such methods will be the choice of \textit{ansatz} for the parameterized quantum circuit. 

Prior work has explored finding the ground state of the Lipkin model via VQE with a Bethe solution-inspired circuit ansatz \cite{PhysRevA.104.022412}, VQE with Unitary Coupled Cluster (UCC) ansatz \cite{chikaoka_quantum_2022}, a Hamiltonian-Learning VQE method \cite{PhysRevC.108.024313} and hybrid classical-quantum algorithms targeting excited states \cite{refId0,hlatshwayo_simulating_2022}.
Exploratory quantum computing calculations of realistic shell model Hamiltonians using VQE have been undertaken for $^6$Li using the Cohen-Kurath interaction \cite{kiss_quantum_2022},  isotopes of He, Be and O with the Cohen-Kurath interaction in the p shell and USD in the sd shell \cite{PhysRevC.105.064308}, a series of oxygen isotopes out to the neutron drip line using USDB and interactions derived from bare NN potentials: Daejeon16, N3LO EFT and JISP16 \cite{sarma_prediction_2023}.  Other work has explored the components of the VQE process for the Agassi model~\cite{perez2022digital} and for future larger quantum shell model calculations~\cite{perez-obiol_nuclear_2023} with the ADAPT-VQE ansatz.
Here, we investigate the Hamiltonian Variational Ansatz~\cite{wecker2015progress} (HVA) as applied to nuclear problems, beginning with the highly simplified and very symmetric Lipkin model, then moving on to the Agassi model. 
% A key feature of the typical quantum computation approach to the nuclear shell model is to propose a wavefunction ansatz, with which it is possible to build in the correlations that in a classical shell model calculation come from diagonalising a matrix in an uncorrelated single-particle basis.  
% The previously cited shell model explorations have used either the generic UCC or the problem specific ADAPT-VQE ans\"atz

A major obstacle for executing variational quantum algorithms at scale are the problems of barren plateaus (BP) and exponential concentration~\cite{mcclean2018barren, cerezo2020cost, holmes2020barren, arrasmith2021equivalence, marrero2020entanglement, wang2020noise}, which exponentially increases the resources required to train a parameterized quantum circuit. These phenomena were initially studied for generic hardware-efficient ansatze~\cite{mcclean2018barren, holmes2021connecting}. Subsequent work studied more specific circuit structures~\cite{schatzki2022theoretical}, and found that encoding symmetries into the ans\"atze has a positive impact of trainability by increasing the variance of gradients ~\cite{larocca2022diagnosing}. Recently this relationship has been formally proven~\cite{ragone2023unified, fontana2023adjoint}. Specifically, it was shown that if the Dynamical Lie Algebra (DLA) is polynomially scaling, the QNN has guaranteed trainability as the cost function variances will decay at worst polynomially with system size.

Techniques for encoding symmetries into the ans\"atze of quantum neural networks, also known as ``geometric quantum machine learning''~\cite{larocca2022group} have been explored for encoding classical data~\cite{meyer2023exploiting, Chang2023Approximately} and condensed matter physics \cite{sauvage2022building, lyu2023symmetry}. Here we study the effect of encoding symmetries into the ans\"atze for the Variational Quantum Eigensolver (VQE) applied to nuclear Hamiltonians. In particular, we study two simplified nuclear models, the Lipkin model and Agassi model, and develop ans\"atze tailored to these models' symmetries. We numerically demonstrate that these symmetrised models, in contrast to a non-structured VQE ansatz, exhibit non-exponentially vanishing gradients and further show the resulting improvements to training. 

While our work here shows that incorporating symmetries provides a simple solution to the BP problem in the context of nuclear problems, this comes at the expense of also making the problem easier to classically simulate. Given the link established in Refs~\cite{ragone2023unified, fontana2023adjoint} between cost variance and DLA scalings, our work  indirectly implies that the Lipkin models and Agassi models have poly DLAs (something which would be challenging to compute directly by other means). It is then known that systems with a small DLA are efficient to simulate classically~\cite{goh2023lie, cerezo2023does}. Thus our work can also be viewed as opening up new quantum-inspired  avenues for classically simulating these models. Alternatively, one can note that more realistic nuclear models will be less symmetric and thus not possible to treat exactly with the fully symmetrised ans\"{a}tze advocated here. We suggest in such contexts our approach will provide a valuable (either classical or quantum) pre-training strategy to obtain a good initialisations for more complex models. 

\section{Lipkin model} \label{sec:lipkin}

The Lipkin model was introduced in \cite{lipkin1965validity} as a simplified model of a nuclear two level system, in which $n$ particles are arranged in two $n$-fold degenerate levels.  The particles interact with a two body interaction which either scatters pairs up or down, or promotes one particle while lowering another. As a simplified nuclear shell model it largely serves as a test case for approximation methods.  The model can also be mapped onto a system of $n$ spins which can point either up or down.    When expressed in this way~\cite{vidal2004entanglement}, the Hamiltonian is written as 
\begin{equation}\label{eq:lipkinham}
    H = -\frac{\lambda}{n}\sum_{i<j} X_i X_j - h \sum_i Z_i ,
\end{equation}
where $X_i$ and $Z_i$ are the Pauli $X$ and $Z$ spin matrices acting on the i$^{th}$ particle, and $\lambda$ and $h$ are strengths of the two terms in the Hamiltonian.
The inverse scaling of the first term with respect to the number of particles, $n$, was introduced in~\cite{vidal2004entanglement} to ensure a finite free energy per spin in the thermodynamic limit. 

The Hamiltonian Variational Ansatz (HVA) was introduced in~\cite{wecker2015progress} as a problem-tailored ansatz template for the Variational Quantum Eigensolver. For a Hamiltonian expressed as a sum of non-commuting terms, $H = \sum_k H_k$, one layer of the corresponding HVA ansatz is $U(\theta) = \prod_k e^{-i \theta_k H_k}$. For close to identity initializations the HVA has substantial loss variances~\cite{park2023hamiltonian}; however, for Hamiltonians with an exponentially growing dynamical Lie algebra, the full landscape has a barren plateau~\cite{fontana2023adjoint, ragone2023unified}.

A HVA layer for the Lipkin model is given by
\begin{align}
    U(\theta) &= \prod_{i=1}^L \left( \text{exp}(-i \sum_{j} \gamma_{i,j}Z_{j}) \ \text{exp}(-i \sum_{j<k} \theta_{i,j,k}X_j X_k) \right) \\
    &= \prod_{i=1}^L \left( \prod_{j} \text{exp}(-i \gamma_{i,j}Z_{j}) \ \prod_{j<k} \text{exp}(-i  \theta_{i,j,k}X_j X_k) \right) .\label{eq:lipkin_ansatz}
\end{align}
The Lipkin model, Eq.~\eqref{eq:lipkinham}, is manifestly symmetric with respect to qubit relabelling, and is therefore permutationally symmetric. %The physical origin of this is due to the lack of distance between the sites in the model, meaning all sites interact equally.
To exploit this property, we can 
% construct an ansatz that also respects this symmetry. Specifically, we use the Hamiltonian Variational ansatz design (HVA).
% To enforce the permutation symmetry,
correlate the parameters of an HVA layer such that $\gamma_{i,j} = \gamma_{i}$ and $\theta_{i,j,k} = \theta_{i}$. For clarity, this condition fixes the rotation angles for each generator to be all the same within one layer, and the ansatz is therefore permutationally invariant.

\begin{figure}[t]
\centering
\includegraphics[width = \columnwidth]{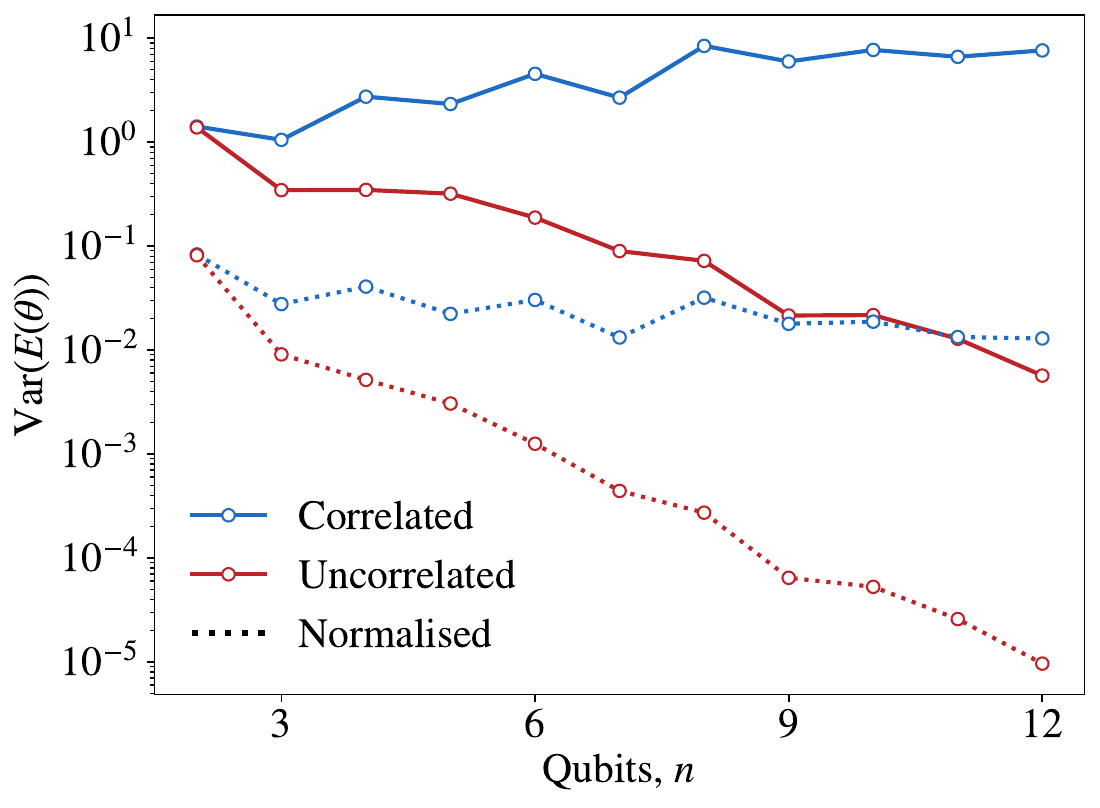}
\vspace{-4mm}
\caption{\textbf{Variance of VQE cost function landscape for the Lipkin model.} Here we show the scaling of the VQE cost function ($E(\theta)=\langle \psi (\theta)|H|\psi(\theta) \rangle$) variances for random initialisations in the parameter space of the ansatz specified in Eq.~\ref{eq:lipkin_ansatz}. We also plot the normalised cost function $\hat{E}(\theta)$ given in Eq.~\ref{eq:normalised_cost} to ensure the cost function is bounded.}
\label{fig:lipkin}
\vspace{-1.5em}
\end{figure}

We investigate the trainability of performing VQE using this ansatz, by computing the average cost function variance across the parameterised landscape.
Fig.~\ref{fig:lipkin} shows a comparison of computing cost function variances, where the cost function for VQE is the energy expectation value. We consider this ansatz in two forms, the first being the permutationally symmetric ansatz with the correlated parameters and the same ansatz structure but un-symmetrised, where all the gate parameters are independent and uncoupled. For both we use a number of layers equal to the number of qubits. For each system size, we evaluate the cost function 32 times, with randomly chosen each parameters from a uniform distribution in the interval $[-2\pi,2\pi]$, which we use to calculate the variance in log-space (i.e., the geometric variance).

The un-symmeterised ansatz exhibits a stereotypical barren plateau, signified by the exponential suppression of energy expectation variances as a function of system size. The magnitude of cost function variances are significantly different for the fully permutationally symmetric ansatz. As opposed to exponentially decaying with system size, we find they in fact increase with system size.
%Specifically, for each ansatz, over 32 random initialisations, we calculate the variance of each parameter gradient, and then take the mean and standard deviation across all gates. We select the ansatz depths to linearly grow with system size.
This could be criticised as being unphysical and an artifact of the increasing size of the eigenspectum of the Hamiltonian. To remove this effect, we also plot the normalised cost function 
\begin{equation}\label{eq:normalised_cost}
    \hat{E}(\theta) = \frac{\langle \psi(\theta)|H|\psi(\theta)\rangle - E_0}{E_{\text{max}} - E_0}
\end{equation}
where $E_0$ and $E_{\text{max}}$ are the minimum and maximum eigenvalues of the Hamiltonian respectively, to ensure $\hat{E}$ is bounded via $0 \leq \hat{E}(\theta) \leq 1$. We still find for the unsymmeterised ansatz the cost function variances exponentially decay with system size, while for the symmetrised ansatz rather than increasing with system size due to the fixed upper bound of 1, they now do not significantly decrease with system size.

While these results give optimism that VQE on the Lipkin model could be performed at scale, it has recently been shown that the ground state energy of permutationally symmetric Hamiltonians can be computed efficiently classically \cite{anschuetz2022efficient}, meaning this is not a path to near-term quantum advantage over classical approaches. It is possible that the ground state outputted by the symmeterised VQE could be used as a input for another quantum algorithm, however this cannot involve evolving the ground-state by a permutationally symmetric Hamiltonian, otherwise this will also be classically simulable \cite{anschuetz2022efficient}.
Although the Lipkin model is potentially too symmetric for its study to gain a quantum advantage, this motivates the search for other Hamiltonians with weaker symmetries that can be also exploited for improved trainability.

%\paragraph*{The Agassi Model}

\begin{figure}[t]
\centering
\includegraphics[width = \columnwidth]{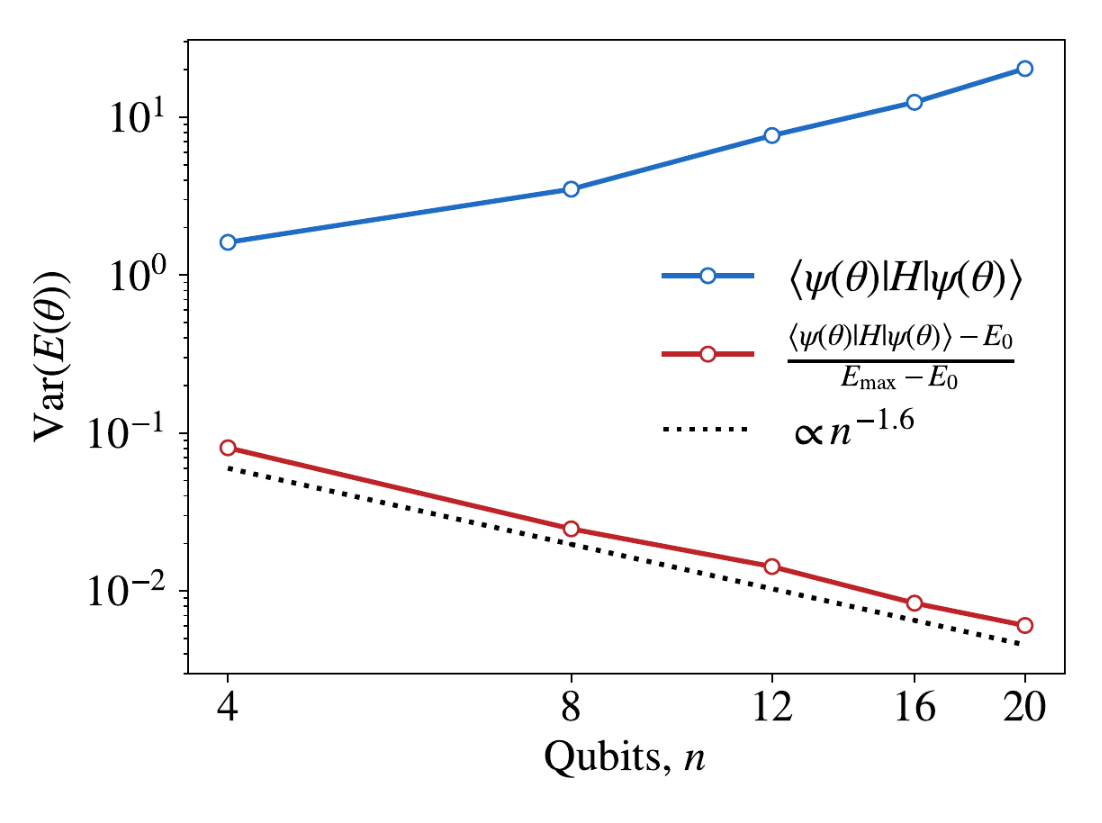}
\vspace{-2mm}
\caption{\textbf{Variance of VQE cost function landscape for the Agassi model.} We plot the variance for both the standard energy expectation value, and the normalised expectation value. We find that the standard energy expectation value grows super-polynomially with system size. The normalised cost function variance decreases polynomially with system size, proportional to $n^{-1.6}$.}
\label{fig:agassi_cost_fct_vars}
\vspace{-1.5em}
\end{figure}

\section{The Agassi Model}

\begin{figure*}[t!]
\centering
\includegraphics[width=\textwidth]{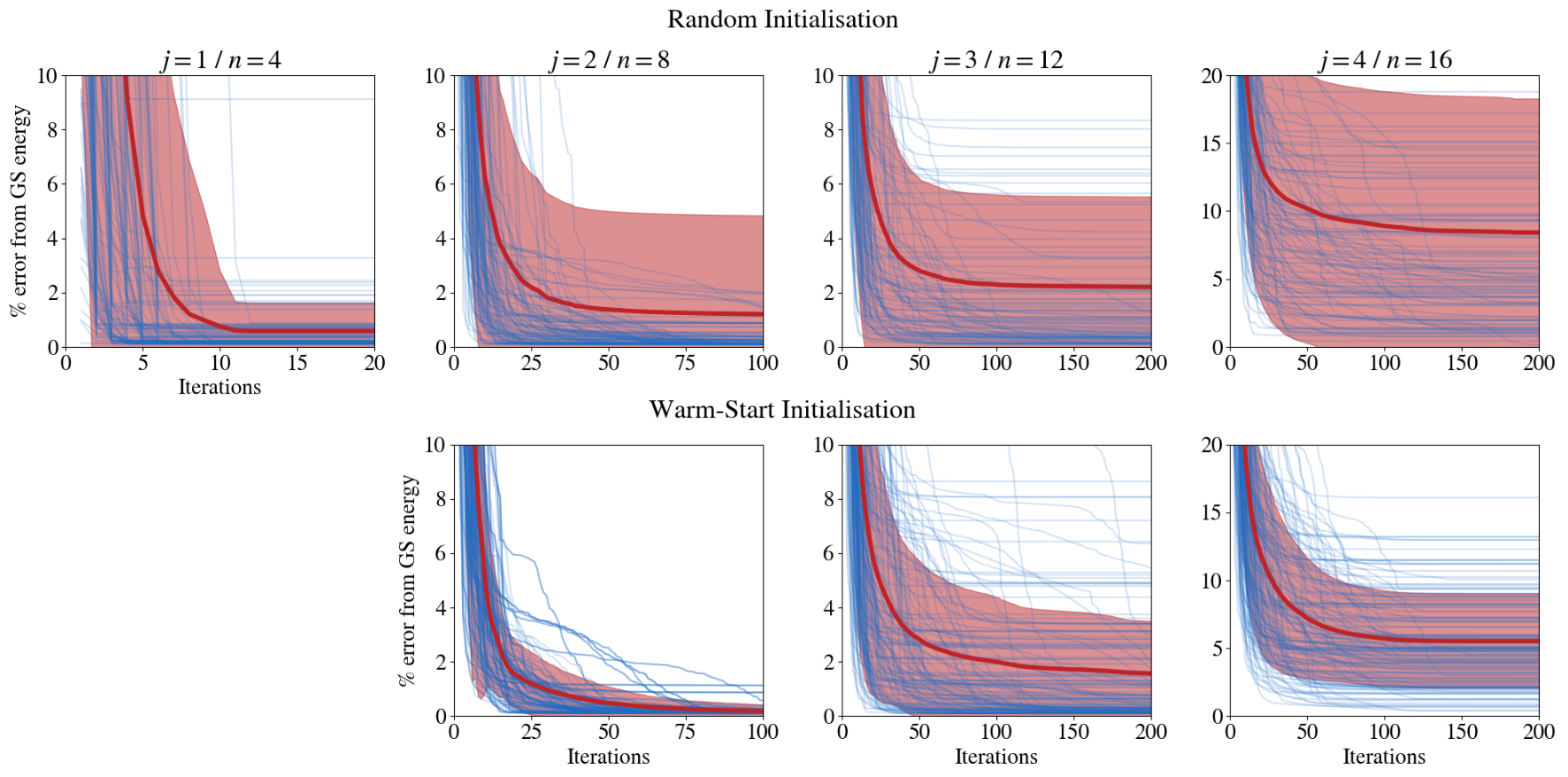}
\vspace{-2mm}
\caption{\textbf{VQE training of the Agassi model.} (Top) The ansatz described in Eq.~\ref{eq:agassi_ansatz} is optimized to prepare the ground state of the Agassi Hamiltonian, for increasing maximum spin $j$, requiring a number of qubits $n=4j$. Plotted in green are the traces of 100 training runs with random initialisations, and the percentage error from the exact ground state energy is plotted as a function of optimisation steps. The red curve and shading represents the mean and standard deviation of this data. (Bottom) Here we use a warm-start initialisation, where we randomly choose one of the solutions from the $j-1^{th}$ warm-started training, with an extra layer of identity gates added.}
\label{fig:agassi_training}
\vspace{-1.5em}
\end{figure*}

The Agassi model is an extension of the Lipkin model, with an extra pairing interaction giving more real physics of a nuclear two-level system. It is typically expressed as
\begin{equation}
    H = \epsilon J^0 - \frac{V}{2} [(J^+)^2+(J^-)^2] - g \sum_{\sigma, \sigma' = -1,1} A_{\sigma}^{\dagger}A_{\sigma'} \, .
    \label{eq:agassi_ham}
\end{equation} 

%These operators are defined in \cite{perez2022digital}, where we also use their qubit ordering convention. To map the fermionic creation/annihilation operators to Pauli spin operators, we use the Jordan-Wigner transform \cite{jordan1993paulische}.
% Here we show the operators in the Hamiltonian defined in term of creation/annihilation operators.
Here $J^0$ and $J^{\pm}$ are defined as 
\begin{align}
    &J^0 = \frac{1}{2} \sum_{m=-j}^j c^{\dagger}_{1,m}c_{1,m} - c^{\dagger}_{-1,m}c_{-1,m} \\
     &J^{+} = \sum_{m=-j}^j c_{1,m}^{\dagger} c_{-1,m} = (J^{-})^{\dagger}.
\end{align} 

% \begin{equation}
%     J^{-} = \sum_{m=-j}^j c_{-1,m}^{\dagger} c_{1,m}
% \end{equation}
Similarly $A^0$ and $A_1$ take the form 
% \begin{equation}
%     A^{\dagger}_1 = \sum_{m=1}^j c^{\dagger}_{1,m}c_{1,-m}^{\dagger}
% \end{equation}
\begin{align}
    &A_{0} = \sum_{m=1}^j c_{1,-m}c_{-1,m} - c_{1,m}c_{-1,-m} \\
    &A_1 = \sum_{m=1}^j c_{1,-m}c_{1,m} \\
    &A_{-1} = \sum_{m=1}^j c_{-1,-m}c_{-1,m}      \, .
\end{align}
% \begin{equation}
%     A_1 = \sum_{m=1}^j c_{1,-m}c_{1,m}
% \end{equation}
% \begin{equation}
%     A_{-1} = \sum_{m=1}^j c_{-1,-m}c_{-1,m}
% \end{equation}
To map the fermionic creation/annihilation operators to Pauli spin operators, we use the Jordan-Wigner transform \cite{jordan1993paulische}.

For performing VQE to produce the ground state of the Agassi model, as done for the Lipkin model, we use the HVA ansatz design, which we find to be efficient in the number of trainable parameters required to prepare the ground state for increasing system sizes.
For convenience, let $H_1 = J^0$, $H_2 = (J^+)^2+(J^-)^2$ and $H_3 = \sum_{\sigma, \sigma' = -1,1} A_{\sigma}^{\dagger}A_{\sigma'}$. %and $H_4 = A_0^{\dagger}A_0$. 
Then our ansatz with $L$-layers is defined by 
\begin{equation}
    U(\theta) = \prod_{i=1}^L \left(\prod_{j=1}^3 \text{exp}(-i \theta_{i,j} H_j) \right)^L
    \label{eq:agassi_ansatz}
\end{equation}
We choose to target the ground state in the half-filling sector.
To do this, we add an extra penalty term to the Hamiltonian to target the correct particle number sector e.g.
\begin{equation}
    H_{\text{Half-Filling}} = H + \beta(\hat{N}_1 + \hat{N}_{-1} - 2j)^2
\end{equation}
for a sufficiently large $\beta$, where 
$
N_\sigma = \sum_{m=-j}^j c^{\dagger}_{\sigma,m}c_{\sigma,m} \, .
$
As the HVA is inspired by adiabatic computing \cite{tilly2022variational}, the initial state is chosen to be the ground state of one of the terms in the Hamiltonian.
A natural starting initial state is $|\psi_0 \rangle = |1\rangle^{\otimes 2j} |0\rangle^{\otimes 2j}$, which is the ground state of the $J^0$ operator and is easily preparable. We use the same qubit ordering as in Ref.~\cite{perez2022digital}, where this state corresponds to the $2j$ particles filling the bottom level of the 2-level system.

%\subsection{Scaling of circuit depth}

%Here we briefly analyse the circuit depth required to implement our ansatz design. Our results shown in Fig XXX heuristically find the ground state can be found with a circuit depth of $\mathcal{O(j)}$ layers for a maximum spin of $j$. The scaling of the circuit depth of these layers is governed by the efficiency of the decomposition of the $n$-body fermionic operator after a Jordan-Wigner transform, into the native gateset of the quantum computer, which we assume can implement up to 2-local gates. This decomposition will require a Suzuki-Trotter decomposition, where for the simplest first-order decomposition has the following error.

%\begin{equation} \label{eq:trotter_error}
%    e^{-i\sum_i^m H_i \theta} = (e^{-i\sum_i^m H_i \frac{\theta}{r}})^r + \mathcal{O}(m^2 t^2 /r^2)
%\end{equation}

%Eq.~\ref{eq:trotter_error} implies that linearly increasing the the parameter repetition parameter $r$, and thus a linear increase in circuit depth, corresponds to a quadratic decay in the error in this Trotter decomposition. Higher order decompositions exist with a smaller Trotter error for a more compact circuit REF Andrew. How this Trotter error affects the ability of the ansatz to prepare the ground state is unclear, but is the subject of future work.

%Finally we consider the circuit depth required for each 

% \section{Results}

\medskip

We first study the trainability of this ansatz and Hamiltonian combination by quantifying the severity of the barren plateau phenomenon. We note the ansatz design described in Eq.~\ref{eq:agassi_ansatz} requires $3j$ parameters for a maximum spin of $j$. For increasing system sizes of the Agassi model, we randomly sample the cost function landscape and compute the variance of the measured energy expectation value. We set the number of layers to equal to $j$ (and therefore linearly increasing with system size). % and each parameter value is randomly sampled 1024 times from the range $[-10, 10]$ (the generators of our gates do not square to the identity, so we do not have the typical $2\pi$ periodicity of Pauli rotation gates). 
Often parameterised gates in variational ans\"atze are exponentiations of single qubit Pauli matrices, and therefore their action with respect to the rotation angle has a period of 4$\pi$, e.g. if $G^2 = 1$, and $R(\theta)=e^{-i \frac{\theta}{2}G} = \text{cos}(\frac{\theta}{2})I - i \ \text{sin}(\frac{\theta}{2})G$, then $R(\theta+4\pi)=R(\theta)$. The gate generators used in Eq.~\ref{eq:agassi_ansatz} do not square to the identity, so  does not have the same $4\pi$ periodicity. Therefore we increase the parameter range uniformly sampled from to $[-10, 10]$.

Fig.~\ref{fig:agassi_cost_fct_vars} shows these results. Similar to the Lipkin results presented in Section~\ref{sec:lipkin}, we again find cost function variances that increase with system size. Plotting the normalised cost function~\footnote{We stress that this normalisation would not be implemented in practise when performing the algorithm on real quantum hardware. Indeed, it would require already knowing the ground state energy. Therefore the growing cost function variances indicate this ansatz is highly trainable, even at scale.} to remove the effect of the increasing size of the spectrum, the variances do now decrease with system size but polynomially scaling as $n^{-1.6}$.

We now demonstrate the ability of this ansatz to produce the ground state of the Agassi model. The top row of Fig.~\ref{fig:agassi_training} shows the results of 100 training runs with random initialisations, for increasing system sizes. Plotted in blue are the traces of individual optimisations, where the percentage error from the exact ground state energy is plotted as a function of optimisation steps. The red curve and shading represents the mean and standard deviation of this data. The ansatz is compact in the number of parameters, as empirically we find we need only a linear circuit depth and number of parameters as a function of system size to prepare the ground state to within a 1\% error. 

We also investigate the benefit gained from warm-starting the training. A feature of the ansatz is that the number of parameters in one layer is independent of system size, as each of the gates are $n$-qubit operators. We can hypothesise there is a similarity between the solutions in parameter space for neighbouring values of $j$, particularly for larger system sizes as the thermodynamic limit is approached. Specifically, for a maximum spin $j$, we randomly sample from the $j$-1 warm-started solutions that achieved a percentage error less than 2\%, and use the output parameter vector for the initialisation. To keep the linearly increasing circuit depth, we also append one layer of gates initialised close to the identity, where the rotation angle is uniformly sampled from the range $[-10^{-4}, 10^{-4}]$. For the optimisation of the parameterised quantum circuit, we use the gradient-based ADAM optimiser~\cite{kingma2015adam}. We note that the parameterised-shift rule~\cite{schuld2019evaluating} commonly used for evaluating gradients in parameterised quantum circuits only applies to gates with generators that have at most two unique eigenvalues, which is not true for the gates in this Hamiltonian Variational Ansatz. More general rules for exact gradient evaluation has been explored in the literature~\cite{izmaylov2021analytic,wierichs2022general,kyriienko2021generalized} that could be used here instead.

It is clear the warm-starting aids the training process, both in reducing the mean and standard deviation of the termination of the optimisation. Specifically, for $j=2,3,4$, upon termination of the optimisation, we find the mean energy value is reduced by 85.12\%, 30.08\% and 34.29\% respectively, and the standard deviations are reduced by 94.05\%, 42.05\% and 63.85\%. We note that in these implementations we did not include the effect of shot noise. Thus, in contrast to our loss variance analysis in Fig.~\ref{fig:agassi_cost_fct_vars} where we demonstrate the loss gradients do not exponentially vanish with system size, here we are predominantly exploring the effect of local minima.

\medskip

\section{Discussion}

% Inspired by recent works analytically studying the trainability of symmetrised ans\"atze for variational quantum algorithms,
Here we presented a practical study of the trainability of symmetrised ans\"atze for two nuclear Hamiltonians: the Lipkin model and the Agassi model. For both Hamiltonians, we employ a variant of the Hamiltonian Variational Ansatz which has been modified to fully respect the symmetries of the target Hamiltonian. Rather than observing decaying cost variances (as would usually be expected), or indeed the exponential decay of a barren plateau, we instead find that the variances increase with system size. This effect can be attributed to the fact that the models' Hamiltonian eigenspectrum range grows with system size. On artificially normalizing the cost to account for this, we observe that the normalized cost variance decays at worst polynomially. Thus we see that the symmeterised ansatz can indeed be used to avoid the barren plateau phenomenon. 

For both the Lipkin and Agassi models we empirically find that a linearly increasing number of trainable gates is sufficient to prepare the ground state and hence our ansatz design seems to be efficient in circuit depth and trainable parameters. Furthermore, we stress that the number of parameters in each layer of the ansatz does not scale with the number of qubits. This observation allowed us to motivate a simple parameter transfer strategy for using solutions from smaller system sizes to provide a warm-start initialisation. Our numerics indicate that this strategy has a positive effect of the success of the optimisation. 

Determining the scaling of the dimension of the DLA is critical parameter for understanding both the trainability of the QNN, and the classical simulability of the target problem, however in practise this a challenging task. While the DLA can be studied numerically, the naive brute-force calculation is costly and scales exponentially with system size, limiting its application~\footnote{see Appendix E of \cite{larocca2022diagnosing} for a discussion of the algorithm}. In some cases the DLA scaling has been determined analytically ~\cite{wiersema2023classification,schatzki2022theoretical}, however this is a challenging task for more general Hamiltonians. As a result, although there is a clear relationship between the DLA and both trainability and classical simulability, in reality the DLA scaling can be difficult to determine in practice.

If the recently derived relationship between the size of the DLA subspace and cost function variances is inverted, our results indicate the Agassi model has a polynomially scaling DLA, implying the system will be efficiently classically simulable by techniques that exploit this property \cite{goh2023lie}. This provokes the use of new quantum-inspired classical algorithms for studying nuclear Hamiltonians.

Moving forward, it would be interesting to study how symmeterized ans\"{a}tze could be applied to ground state computations of more complex nuclear models. In particular, it is natural to investigate whether a symmeterised solution to one of the simpler models studied here might provide a good initialization for future larger nuclear shell model calculations~\cite{perez-obiol_nuclear_2023}. A natural test case could be along an isotopic chain for a magic number of protons, e.g. in oxygen as neutrons are added across and beyond the $sd$ shell \cite{brown_oxygen_2017}.%Natural test cases here include XXXX and XXXX. 

%Our results, in conjunction with the recently derived formula for cost function variances, indicate the Agassi model has a polynomially scaling Dynamical Lie Algebra, implying the system will be efficiently classically simulable by techniques that exploit this property \cite{goh2023lie}. 

%Future work could prove this analytically, and investigate whether extensions of this model are also classically simulable.

%\zo{I think it would be nice to add some discussion here saying that the poly variance scaling implies that the algebra should be polynomial even though this is hard to verify in practise. And that spinning it around this might imply that this model is classically simulable. Thus this could open new quantum inspired classical algorithms for solving this problem.}

%\zo{I would then probably leave the paper on that cliff hanger.}

%\zo{BUT if you don't like that could instead consider a more complex model again with even less symmetry?}

%\zo{However I'd lean towards saving that for a future project}

\begin{acknowledgments}
We thank Martin Larocca for interesting discussions of Dynamical Lie Algebra. ZH acknowledges support from the Sandoz Family Foundation-Monique de Meuron program for Academic Promotion.  PS acknowledges support from the UK STFC under grants ST/W006472/1 and ST/V001108/1, and support from AWE under the William Penney Fellowship scheme. UK Ministry of Defence $\copyright$Crown owned copyright 2024/AWE
\end{acknowledgments}

\bibliography{refs, quantum}

\clearpage 

\end{document}